\documentstyle[12pt,psfig]{article}

\catcode`\@=11
\long\def\@makefntext#1{
\protect\noindent \hbox to 3.2pt {\hskip-.9pt  
$^{{\ninerm\@thefnmark}}$\hfil}#1\hfill}                %CAN BE USED 

 \def\@makefnmark{\hbox to 0pt{$^{\@thefnmark}$\hss}}  %ORIGINAL 
        
\def\ps@myheadings{\let\@mkboth\@gobbletwo
\def\@oddhead{\hbox{}
\rightmark\hfil\ninerm\thepage}   
\def\@oddfoot{}\def\@evenhead{\ninerm\thepage\hfil
\leftmark\hbox{}}\def\@evenfoot{}
\def\sectionmark##1{}\def\subsectionmark##1{}}

\newcounter{sectionc}\newcounter{subsectionc}\newcounter{subsubsectionc}
\renewcommand{\section}[1] {\vspace{0.6cm}\addtocounter{sectionc}{1} 
\setcounter{subsectionc}{0}\setcounter{subsubsectionc}{0}\noindent 
        {\bf\thesectionc. #1}\par\vspace{0.4cm}}
\renewcommand{\subsection}[1] {\vspace{0.6cm}\addtocounter{subsectionc}{1} 
        \setcounter{subsubsectionc}{0}\noindent 
        {\it\thesectionc.\thesubsectionc. #1}\par\vspace{0.4cm}}
\renewcommand{\subsubsection}[1] {\vspace{0.6cm}\addtocounter{subsubsectionc}{1}
        \noindent {\rm\thesectionc.\thesubsectionc.\thesubsubsectionc. 
        #1}\par\vspace{0.4cm}}

\newcounter{appendixc}
\newcounter{subappendixc}[appendixc]
\newcounter{subsubappendixc}[subappendixc]

\renewcommand{\appendix}[1] {\vspace{0.6cm}
        \refstepcounter{appendixc}
        \setcounter{figure}{0}
        \setcounter{table}{0}
        \setcounter{equation}{0}
        \renewcommand{\thefigure}{\Alph{appendixc}.\arabic{figure}}
        \renewcommand{\thetable}{\Alph{appendixc}.\arabic{table}}
        \renewcommand{\theappendixc}{\Alph{appendixc}}
        \renewcommand{\theequation}{\Alph{appendixc}.\arabic{equation}}
        \noindent{\bf Appendix \theappendixc #1}\par\vspace{0.4cm}}

\def\abstracts#1{{
        \centering{\begin{minipage}{30pc}\tenrm\baselineskip=12pt\noindent
        \centerline{\tenrm ABSTRACT}\vspace{0.3cm}
        \parindent=0pt #1
        \end{minipage}}\par}} 

\renewenvironment{thebibliography}[1]
        {\begin{list}{\arabic{enumi}.}
        {\usecounter{enumi}\setlength{\parsep}{0pt}
\setlength{\leftmargin 1.25cm}{\rightmargin 0pt}
         \setlength{\itemsep}{0pt} \settowidth
        {\labelwidth}{#1.}\sloppy}}{\end{list}}

\newcounter{itemlistc}
\newcounter{romanlistc}
\newcounter{alphlistc}
\newcounter{arabiclistc}

\newcommand{\fcaption}[1]{
        \refstepcounter{figure}
        \setbox\@tempboxa = \hbox{\tenrm Fig.~\thefigure. #1}
        \ifdim \wd\@tempboxa > 6in
           {\begin{center}
        \parbox{6in}{\tenrm\baselineskip=12pt Fig.~\thefigure. #1}
            \end{center}}
        \else
             {\begin{center}
             {\tenrm Fig.~\thefigure. #1}
              \end{center}}
        \fi}

\newcommand{\tcaption}[1]{
        \refstepcounter{table}
        \setbox\@tempboxa = \hbox{\tenrm Table~\thetable. #1}
        \ifdim \wd\@tempboxa > 6in
           {\begin{center}
        \parbox{6in}{\tenrm\baselineskip=12pt Table~\thetable. #1}
            \end{center}}
        \else
             {\begin{center}
             {\tenrm Table~\thetable. #1}
              \end{center}}
        \fi}

\def\Journal#1#2#3#4{{#1} {\bf #2}, #3 (#4)}

\def\NCA{{\em Nuovo Cimento} A}
\def\NPB{{\em Nucl. Phys.} B}
\def\PLB{{\em Phys. Lett.}  B}
\def\PRL{\em Phys. Rev. Lett.}
\def\PRD{{\em Phys. Rev.} D}
\def\ZPC{{\em Z. Phys.} C}

\def\NPA{{\em Nucl. Phys.} A}
\def\JPG{{\em J. Phys.} G}
\def\nn{\nonumber}

\def\vk{{\bf k}_{\perp}}
\def\al{\alpha_s}
\def\be{\begin{equation}}
\def\ee{\end{equation}}
\def\bea{\begin{eqnarray}}
\def\eea{\end{eqnarray}}
\def\rd{{\rm d}}
\def\kev{\,{\rm keV}}
\def\mev{\,{\rm MeV}}
\def\gev{\,{\rm GeV}}
\newcommand{\da}{distribution amplitude}
\newcommand{\das}{distribution amplitudes}

\def\fnt#1#2{\footnotetext{\kern-.3em
        {$^{\mbox{\sevenrm #1}}$}{#2}}}

 1
 1
 1

\font\tenbf=cmbx10
\font\tenrm=cmr10
\font\tenit=cmti10

\font\ninerm=cmr9

\begin{document}
\thispagestyle{empty}
\hspace*{9.0cm}                        WU-B 97-1\\
\hspace*{9.5 cm}                      January 1997\\            
\\
\vspace*{1cm}
\begin{center}
{\Large\bf THE $\pi\gamma$ TRANSITION FORM FACTOR\\
        AND ITS IMPACT ON CHARMONIUM DECAYS INTO TWO PIONS
\footnote {Invited talk presented at the Third Workshop
on Diquarks, Torino 
(October 1996) }} \\
\vspace*{1.0 cm}
\end{center}
\begin{center}
{\large P. Kroll}\\
\vspace*{0.5 cm}
Fachbereich Physik, Universit\"{a}t Wuppertal, \\
D-42097 Wuppertal, Germany\\[0.3 cm]
\end{center}
\newpage
\setcounter{page}{1}
\centerline{\tenbf THE $\pi\gamma$ TRANSITION FORM FACTOR} 
\baselineskip=22pt
\centerline{\tenbf AND ITS IMPACT ON CHARMONIUM DECAYS INTO TWO PIONS}
\vspace*{0.2cm}
\centerline{\tenrm PETER KROLL \footnote{ e-mail: 
kroll@theorie.physik.uni-wuppertal.de\\
Supported in part by the TMR Network ERB 4061 PL 95 0115.}}
\baselineskip=13pt
\centerline{\tenit Fachbereich Physik, Universit\"at Wuppertal, }
\baselineskip=12pt
\centerline{\tenit D-42097 Wuppertal, Germany}
\vspace*{0.5cm}
\abstracts{The analysis of the $\pi\gamma$ transition form factor
provides severe constraints on the pion's wave function. This
information is used to examine charmonium decays into two pions critically. 
It will be argued that the standard perturbative QCD analysis of these
reactions fails, i.~e.~the need for additional contributions can convincingly
be demonstrated. Colour-octet admixtures to the charmonium states are
proposed as a possible dynamical mechanisms to solve the
puzzle. Consequences of the $\pi\gamma$ analysis for the
electromagnetic form factor of the pion are also discussed. }
\vfil
\vspace*{0.8cm}
\rm\baselineskip=14pt
%%%%%%%%%%%%%%%%%%%%%%%%%%%%%%%%%%%%%%%%%%%%%%%%%%%%%%%%%%%%%%%%%%%%%%%%%%%
\section{Introduction}
\label{sec:intro}
%%%%%%%%%%%%%%%%%%%%%%%%%%%%%%%%%%%%%%%%%%%%%%%%%%%%%%%%%%%%%%%%%%%%%%%%%%%
\vspace*{-0.3cm}
At large momentum transfer the hard scattering approach (HSA) \cite{lep:80} 
provides a scheme to calculate exclusive processes. Observables are 
described as convolutions of hadronic wave functions which embody 
soft non-perturbative physics, and hard scattering amplitudes $T_H$ 
to be calculated from perturbative QCD. In most cases only the 
contribution from the lowest-order pQCD approach in the collinear 
approximation using valence Fock states only (termed the standard HSA)
has been worked out. Applications of the standard HSA to space-like
exclusive reactions, as for instance the magnetic form factor of the
nucleon, the pion form factor or Compton scattering off protons
revealed that the results are only in fair agreement with experiment
if hadronic wave functions are used that are strongly concentrated in
the end-point regions where one of the quark momentum fractions, $x$, 
tends to zero. As has been pointed out by several authors 
(e.g.\ \cite{isg:89,rad:91}), the results obtained from such wave 
functions are dominated by contributions from the end-point regions 
where perturbative QCD cannot readily be applied. Hence, despite the 
agreement with experiment, the predictions of the standard HSA are 
theoretically inconsistent for such wave functions. It should also be 
stressed that the large momentum transfer behaviour of the 
helicity-flip controlled Pauli form factor of the proton remains 
unexplained within the standard HSA. 

Applications of the HSA to time-like exclusive processes fail in most
cases (e.g.\ $G_M$, $F_{\pi}$, $\gamma\gamma\to p\bar{p}$). 
The predictions for the integrated  $\gamma\gamma\to \pi\pi$
cross-section ($|\cos{\theta}| \leq 0.6$) are in fair
agreement with the data whereas the predictions for the angular
distribution fails. Exclusive charmonium decays constitute another
class of time-like reactions. If the end-point region concentrated wave
functions are employed again, the standard HSA provides results 
in fair agreement with the data in many cases. 
It should be noted that in most calculations of exclusive
charmonium decays \cite{dun:80} $\al$ values of the 
order of $0.2 - 0.3$ are employed. Such values do not match with $\al$ 
evaluated at the charm quark mass, the characteristic scale for these 
decays ($\al (m_c=1.5 \gev) = 0.37$ in one-loop approximation with 
$\Lambda_{QCD}=200 \mev$). Since high powers of $\al$ are involved in 
charmonium decays a large factor of uncertainty is hidden in the predictions.   

Constraining the pion wave function \cite{jak:96,kro:96} from
the recent precise data on the $\pi\gamma$ transition form factor
\cite{cleo:95}, one observes an order-of-magnitude 
discrepancy between data and HSA predictions for charmonium decays
into two pions. In \cite{bol:96} contributions from the 
$c\bar{c}g$ Fock state are suggested as the solution of this puzzle.
I am going to discuss these topics in my talk. Also I shall discuss
the large momentum transfer behaviour of the pion form factor in the 
light of the new information on the pion's wave function.

%%%%%%%%%%%%%%%%%%%%%%%%%%%%%%%%%%%%%%%%%%%%%%%%%%%%%%%%%%%%%%%%%%%%%%%%%%%
%\vspace*{-0.3cm}
%%%%%%%%%%%%%%%%%%%%%%%%%%%%%%%%%%%%%%%%%%%%%%%%%%%%%%%%%%%%%%%%%%%%%%%%
\section{The $\pi$-$\gamma$ transition form factor}
\label{sec:pigaff}
\begin{figure}[t]
%\vspace*{-2cm}
\[
    \psfig{figure=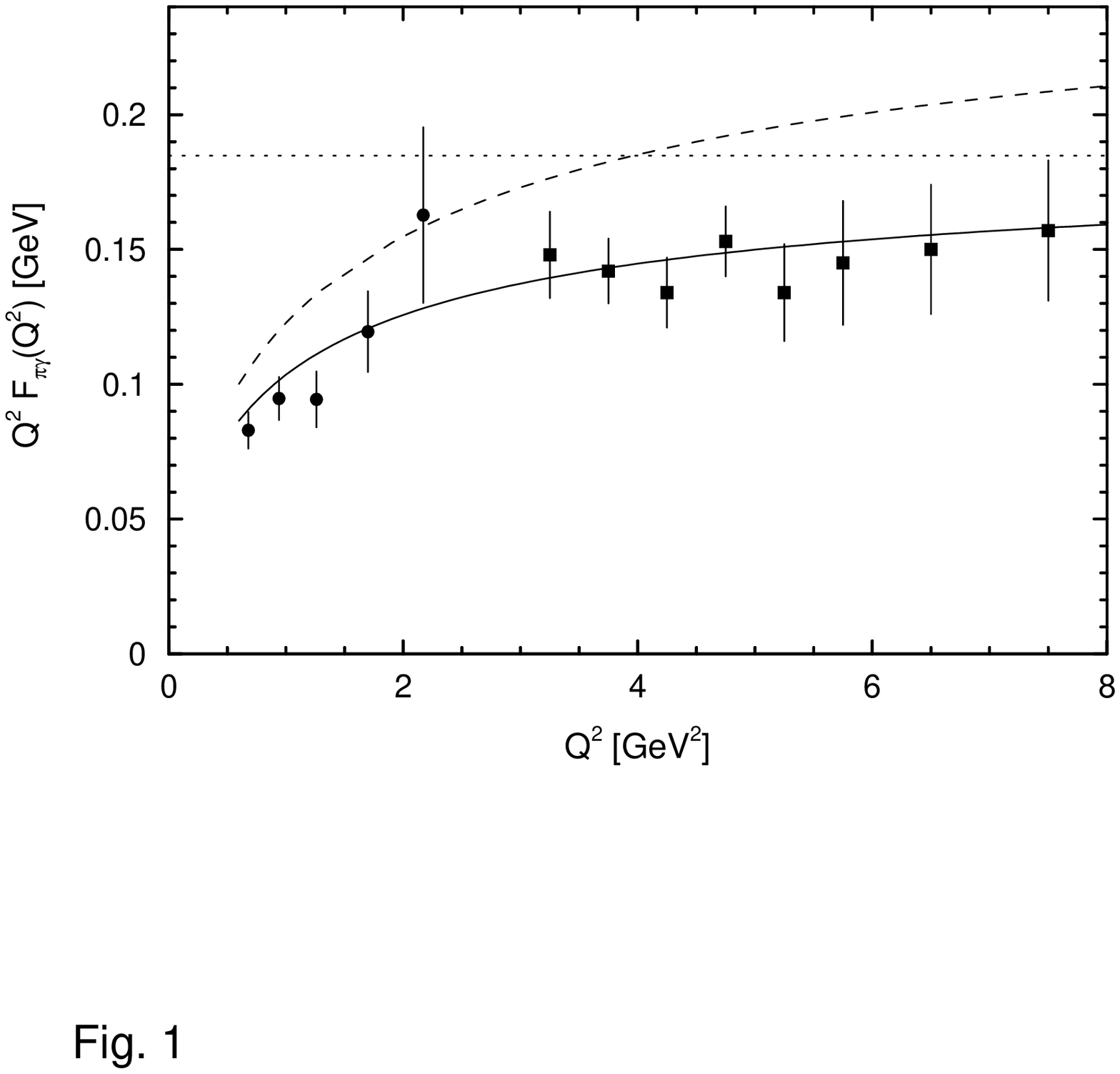,%
        bbllx=20pt,bblly=270pt,bburx=570pt,bbury=660pt,%  
%       bbllx=2.5cm,bblly=18cm,bburx=18cm,bbury=25cm,%
        width=8cm,clip=}
%       width=6cm,clip=}
\]
\vspace*{-1.0cm}
\caption[dummy2]{The scaled $\pi\gamma$ transition form factor vs.\
$Q^2$. The solid (dashed) line represents the results obtained with 
the modified HSA using the asymptotic (Chernyak-Zhitnitsky) wave
function.  The evolution of the Chernyak-Zhitnitsky wave function is 
taken into account. The dotted line represents the limiting behaviour 
$\sqrt 2f_\pi$. Data are taken from \cite{cleo:95,cello:91}.}
\label{fig:pigaff}
\end{figure}

%%%%%%%%%%%%%%%%%%%%%%%%%%%%%%%%%%%%%%%%%%%%%%%%%%%%%%%%%%%%%%%%%%%%%%%%
\vspace*{-0.3cm}
The apparent success of the end-point concentrated wave
functions, in spite of the theoretical inconsistencies, prevented progress
in understanding hard exclusive reactions for some time. Recently,
with the advent of the CLEO data on the $\pi\gamma$ transition form 
factor $F_{\pi\gamma}$ \cite{cleo:95}, the situation has changed. 
The leading twist result for that form factor\footnote{
The pion mass as well as the light current quark masses are neglected throughout.}, 
including $\al$-corrections, reads \cite{lep:80}
\be
\label{leadtwisteq}
F_{\pi\gamma}(Q^2) = \frac{\sqrt 2}{3}\,\langle
                 x^{-1}\rangle\frac{f_\pi}{Q^2}\;
             [\,1+\frac{\alpha_s(\mu_R)}{2\pi}
               K_{\pi\gamma}(Q^2,\mu_R) + {\cal O}(\alpha_s^2)\,]. 
\ee  
$f_{\pi}$ is the usual pion decay constant (130.7 \mev) and $\mu_R$ 
represents the renormalization scale. The function $K_{\pi\gamma}$ has
been calculated by Braaten \cite{bra:83} in the $\overline{MS}$ scheme. 
$\langle x^{-1}\rangle$ is the $1/x$ moment of the
pion distribution amplitude, $\phi$, which represents the
light-cone wave function of the pion integrated over transverse
quark momenta, $\vk$, up to a factorization scale, $\mu_F$, of order $Q$. 
The \da\ can be expanded upon Gegenbauer polynomials, $C_n^{3/2}$, the 
eigenfunctions of the evolution kernel for mesons \cite{lep:80}
\be
\label{evoleq}
\phi_{\pi}(x,\mu_F)=\phi_{AS}(x)\left[1+ 
    \sum^\infty_{n=2,4,...}B_n(\mu_0)\left(
\frac{\al\left(\mu_F\right)}{\al\left(\mu_0\right)}\right)
^{\gamma_n}\,C_n^{3/2}(2x-1)\right] 
\ee
where the asymptotic \da\ is $\phi_{AS}(x)=6x(1-x)$. The $1/x$
moment of the \da\ reads
\be
\langle x^{-1}\rangle=3\left[1+\sum^\infty_{n=2,4,...}B_n(\mu_0)
           \left(\frac{\alpha_s(\mu_F)}{\alpha_s(\mu_0)}\right)^{\gamma_n}\right] 
          = 3\left[1+\sum^\infty_{n=2,4,...} B_n(\mu_F)\right].
\label{leadmomeq}
\ee
The process-in\-depen\-dent ex\-pan\-sion coefficients $B_n$ embody
the soft physics; they are not calculable at present. $\mu_0$ is a 
typical hadronic scale, actually $\mu_0=0.5 \gev$. Since the anomalous
dimensions, $\gamma_n$, are positive fractional numbers increasing
with $n$ (e.g.\ $\gamma_2=50/81$) any \da\ evolves into the asymptotic
\da\ for $\ln{Q^2}\to\infty$; higher order terms are gradually suppressed. 
Hence, the limiting behaviour of the transition form factor is
\be
\label{asy}
F_{\pi\gamma} \longrightarrow \sqrt{2} f_{\pi} /Q^2
\ee
which is a parameter-free QCD prediction \cite{wal:74}. As comparison 
with the CLEO data \cite{cleo:95} reveals, the limiting
behaviour is approached from below. At 8 \gev$^2$ the data only
deviate by about $15 \%$ from (\ref{asy}) (see Fig.\ref{fig:pigaff}). 
In order to give a quantitative estimate of the allowed deviations from the
asymptotic \da\, one may assume that $B_2$ is the only non-zero
expansion coefficient in (\ref{evoleq}). The truncated series suffices to
parametrize small deviations. Moreover, it has the advantage of
interpolating smoothly between the asymptotic \da\ and the frequently
used Chernyak-Zhitnitsky \da\ \cite{che:82} ($B_2=2/3$; 
$C_2^{3/2}(\xi)=3/2(5\xi^2-1)$). For large momentum transfer the
assumption on the expansion coefficients is
justified by the properties of the anomalous dimensions $\gamma_n$. 

In \cite{kro:96} it is shown that the leading twist, lowest order pQCD
result (\ref{leadtwisteq})  nicely fits the CLEO data for 
$B_2^{LO}(\mu_0)=-0.39\, \pm 0.05$. Using Braaten's result for
$K_{\pi\gamma}$ \cite{bra:83} that, choosing $\mu_F=\mu_R=Q$, reads
\be
\label{corr}
K_{\pi\gamma} = -\frac{10}{3}\;\frac{1-59/72\,B_2 (Q^2)}
                          {1+B_2 (Q^2)}\, ,   
\ee
one finds $B_2^{NLO}(\mu_0)=-0.17\, \pm 0.05$ from a fit to the CLEO 
data. Braaten's analysis is however incomplete in so far as only the 
$\alpha_s$ corrections to the hard scattering amplitude have been 
considered but the corresponding corrections to the kernel of the 
evolution equation for the pion's distribution amplitude were ignored. 
As has been shown by M\"uller \cite{mue:95} recently next-to-leading 
order evolution provides logarithmic modifications in the end-point 
regions for any distribution amplitude, i.e.\ for the asymptotic one too.
An estimate however reveals that the modifications of the evolution
behaviour in next-lo-leading order are very small and can safely be
neglected.

To summarize the $F_{\pi\gamma}$ form factor requires a \da\ in a
leading twist analysis that is narrower than the asymptotic one 
in the momentum transfer region of a few 
\gev$^2$. The Chernyak-Zhitnitsky \da\ is in clear
conflict with the data and should, therefore, be discarded\footnote{
In \cite{BHL:83,cao:96} a modification of the pion wave function is 
proposed where the \da\ is multiplied by the exponential 
$\exp\left[-m_q^2a_{\pi}^2/x(1-x)\right]$. The parameter $m_q$ represents a 
constituent quark mass of, say, $330\mev$. Since the exponential
substantially deviates from unity only in the end-point regions it
leads to a strong additional suppression in the case of the
Chernyak-Zhitnitsky \da\ ($\langle x^{-1}\rangle$ changes from a value
of 5 to 3.71 at the scale $\mu_0$). For narrow \das\
($B_2\leq 0$), on the other hand, the exponential has only a minor 
bearing on the results for $F_{\pi\gamma}$.}.

Recently a modified HSA has been proposed by Botts, Li and Sterman
\cite{bot:89} in which transverse degrees of freedom as well as
Sudakov suppressions are taken into account. This approach has the
advantage of strongly suppressed end-point regions. Hence, the
perturbative contributions can be calculated self-consistently in the
sense that the bulk of the perturbative contribution is accumulated in
regions of reasonably small values of the strong coupling constant.
It is to be stressed that the effects of the transverse degrees of freedom taken 
into account in the modified HSA represent soft contributions of 
higher-twist type. Still, modified HSA calculations are restricted to the 
dominant (valence) Fock state. Another advantage of the modified HSA
is that the renormalization scale can be chosen in such a way that
large logs from higher order perturbation theory are eliminated. Such
a choice of the renormalization scale are accompanied by $\al$ singularities in
the end-point regions which are, however, compensated by the Sudakov
factor in the modified HSA. Singularities produced by the evolution of
the wave function are also cancelled by the Sudakov factor. 

Adapting the modified HSA to the case of $\pi\gamma$ transitions, 
one can write the corresponding form factor as \cite{jak:96,kro:96}
\be
F_{\pi\gamma}(Q^2)=\int \rd x\,\frac{\rd^2{\bf b}}{4\pi}
\hat\Psi_{\pi}(x,-{\bf b},\mu_F)\,
\hat T_H\left(x,{\bf b},Q\right)\,\exp\left[-S\left(x,b,Q\right)\right]
\label{fpgeq}
\ee
up to $\cal {O}$($\al$, $k^2_{\perp}/Q^2$) corrections. 
${\bf b}$ is the quark-antiquark separation and is
canonically conjugated to the usual transverse momentum ${\bf k}_{\perp}$.
The use of the transverse configuration space is mandatory because the
Sudakov exponent $S$ is only known in that space \cite{bot:89}.
The Sudakov exponent comprises those gluonic radiative corrections
not taken into account in the evolution of the wave function.
$\hat T_H$ is the Fourier transform of the lowest order momentum space hard
scattering amplitude. It reads
\be
\hat T_H\left(x,{\bf b},Q\right)=\frac{2}{\sqrt 3\pi}K_0
\left(\sqrt{1-x}\,Q\,b\right)
\label{hattheq}
\ee
where $K_0$ is the modified Bessel function of order zero. 
Due to the properties of the Sudakov exponent any 
contribution is damped asymptotically, i.e. for $\ln (Q^2/\mu_0^2)\to\infty$, 
except those from configurations with small quark-antiquark separations and, 
as can be shown, the limiting behaviour (\ref{asy}) emerges.
$b$ plays the role of an infrared cut-off; it sets up the interface between 
non-perturbative soft gluon contributions - still contained in the hadronic 
wave function - and perturbative soft gluon contributions accounted for by the 
Sudakov factor. Hence, the factorization scale $\mu_F$ is to be taken as $1/b$.

Finally, $\hat\Psi_{\pi}$ is the Fourier transform of the momentum space 
(light-cone) wave function of the pion for which a Gaussian
$\vk$-dependence is employed
\be
\label{gaussian}
\Psi_{\pi}\left(x,\vk;\mu_F\right) = \frac{f_\pi}{2\sqrt 6}\,
                                           \phi_{\pi}(x,\mu_F)
            N\exp{\left(-a^2_{\pi}(\mu_F)\frac{k_{\perp}^2}{x(1-x)}\right)}. 
\ee
Here $N=16\pi^2 a^2_{\pi}/(x(1-x))$ and, for a \da\ with $B_n=0$ for 
$n\geq 4$, $a_{\pi}=1/(\pi f_{\pi} \sqrt{8(1+B_2)}$. The 
$\pi^0\to \gamma\gamma$ constraint \cite{BHL:83} is automatically 
satisfied for that choice of the transverse size parameter $a_{\pi}$. 
$\Psi_{\pi}$ represents a soft wave function, i.e.\ a full wave 
function with its pertubative tail removed from it. For $B_2=0$ the 
wave function (\ref{gaussian}) leads to a valence Fock state
probability of 0.25 and a r.m.s.\ radius of $0.42$ fm. Using the 
wave function (\ref{gaussian}) in a modified HSA calculation, 
one finds excellent agreement with the CLEO \cite{cleo:95} and 
CELLO \cite{cello:91} data above $Q^2 \simeq 1\;\gev^2$ for 
$B_2(\mu_0) =-0.006 \pm 0.014$ \cite{jak:96,kro:96} (see 
Fig.\ref{fig:pigaff}). Hence, the asymptotic wave function, 
i.e.\ the asymptotic \da\ combined with the Gaussian $\vk$-dependence,
works very well if the modified HSA is used.

A similar analysis of the $\eta\gamma$ and the $\eta'\gamma$
transition form factors has been carried through in \cite{jak:96}. It
is important thereby to take into account mass corrections and the
$\eta - \eta'$ mixing. The results of that analysis are in excellent
agreement with the available data including the recent CLEO data
\cite{cleo:95}. The values of the $\eta - \eta'$ mixing angle and the
decay constants are calculated in \cite{jak:96} to be
$\theta_P=-18^{\circ} \pm 2^{\circ}$, $f_{\eta}= 175 \pm 10 \mev$
and $f_{\eta'}= 95\pm 6 \mev$, respectively. The $\eta_c\gamma$
transition form factor can be analyzed in the same manner. Estimates of
that form factor can be found in \cite{aur:96}.

With the advent of the CLEO data the $\pi\gamma$ transition form
factor attracted much interest and, besides \cite{jak:96,kro:96},  
many papers have been devoted to its analysis elucidating various 
aspects of it \cite{cao:96,ong:95,rad:95,ani:96,ans:95}.
Particularly interesting is the generalization of (\ref{leadtwisteq})
to the case of two virtual photons. In the standard HSA and again
with $B_n=0$ for $n\geq 4$, the $\pi\gamma^*$ transition form factor
reads \cite{bra:83} 
\bea 
F_{\pi\gamma^*}(Q^2,\omega)&=& \sqrt{2}\frac{f_{\pi}}{Q^2} \frac{1}{(1-\omega)^3}
                \Big\{ [ 1 - \omega^2 + 2\omega \ln{\omega}] \nn\\
            &&  \times [1 + \frac{1+28\omega+\omega^2}{(1-\omega)^2} B_2(\mu_F)]
                 + 10 \omega \ln{\omega} B_2(\mu_F)\Big \}
\label{two}
\eea
where $\omega=Q'^{2}/Q^2$. The larger one of the two photon
virtualities is denoted by $Q^2$, the smaller one by $Q'^2$. 
The factorization scale may be chosen as $\mu_F= Q \sqrt{1+\omega}$.
$\al$-corrections to $F_{\pi\gamma^*}$ can be found in
\cite{bra:83} and an estimate of power corrections in \cite{gor:89}. 
The treatment of $\pi\gamma^*$ transitions within the
the modified HSA is straightforward generalization of (\ref{fpgeq}) \cite{ong:95}.

Interestingly, the $F_{\pi\gamma^*}$ form factor still behaves as
$Q^{-2}$ at large $Q^2$. This is to be contrasted with the 
$Q^{-2}Q'^{-2}$ behaviour of the vector meson dominance model 
\cite{kes:93}. In the limes $\omega\to 1$ (\ref{two}) simplifies to 
\be
F_{\pi\gamma^*}=\frac{\sqrt{2}}{3}\frac{f_{\pi}}{Q^2} \left [
                 1 + \frac{1}{2}(1 - \omega) (1 - 12 B_2(\mu_F)) \right ].
\label{kappa}
\ee
The limiting behaviour of the form factor for $\omega=1$ which 
is strictly independent on the form of the \da, has also been derived 
from QCD sum rules \cite{nov:84}. In \cite{ans:95} the triangle 
diagram is analyzed with the most general form of the $\pi q \bar{q}$ 
vertex. The result obtained for $F_{\pi\gamma^*}$ in that paper is 
similar to (\ref{two}) provided $B_2$ is put to zero in (\ref{two}). 
The differences between the two results are strongest at $\omega=0$ 
(about $9\%$) while both the results coincide at $\omega=1$.

%%%%%%%%%%%%%%%%%%%%%%%%%%%%%%%%%%%%%%%%%%%%%%%%%%%%%%%%%%%%%%%%%%%%
\section{Pionic decays of charmonium}  
\label{sec:charmonium}
%%%%%%%%%%%%%%%%%%%%%%%%%%%%%%%%%%%%%%%%%%%%%%%%%%%%%%%%%%%%%%%%%%%%
\vspace*{-0.3cm}
In view of the results for $F_{\pi\gamma}$ a fresh analysis of the
decays $\chi_{cJ}\to \pi\pi$ is in order. Using the information on 
the $\pi$ wave function obtained from the analysis of $F_{\pi\gamma}$,
one finds the following values for the partial widths 
\be
\label{shsa}
\Gamma (\chi_{c0(2)}\to\pi^+\pi^-)\, =\, 0.872\; (0.011)\, \kev 
\ee
within the standard HSA \cite{bol:96}. As usual the renormalization 
and the factorization scales are identified in that calculation and 
put equal to the $c$-quark mass. The parameter describing the 
$\chi_{cJ}$ state is the derivative $R'_P(0)$ of the
non-relativistic $c\bar{c}$ wave function at the origin (in coordinate
space) appropriate for the dominant Fock state of the $\chi_{cJ}$, a
$c\bar{c}$ pair in a colour-singlet state with quantum numbers
${}^{2S+1}L_J={}^3P_J$. $m_c=1.5\;\gev$  and, of course, the leading
order standard HSA value -0.39 for $B_2(\mu_0)$ are
chosen as well as $R'_P(0)=0.22\gev^{5/2}$ which is consistent with 
a global fit of charmonium parameters \cite{man:95} as well as with results
for charmonium radii from potential models \cite{buc:81}.
 
In \cite{bol:96} the modified HSA is also used to calculate the
$\chi_{cJ}\to \pi\pi$ decay widths. Taking  $B_2=0$
and the other parameters as quoted above, one finds
\be
\label{mhsa}
\Gamma (\chi_{c0(2)}\to\pi^+\pi^-)\, =\, 8.22\; (0.41)\, \kev. 
\ee
For comparison the experimental data as quoted in \cite{pdg} and
reported in a recent paper of the BES collaboration \cite{bes} are
\bea
\label{dat}
\Gamma (\chi_{c0}\to\pi^+\pi^-)&=& 105\; \pm 30\phantom{.3} \kev \;({\rm PDG}),\nn \\ 
                               && 62.3\pm 17.3 \kev \;({\rm BES}), \nn \\ 
\Gamma (\chi_{c2}\to\pi^+\pi^-)& =& 3.8\;\;\pm 2.0\phantom{3}
                                                      \kev \;({\rm PDG}),\nn \\
                               &&  3.04 \pm 0.73 \kev\; ({\rm BES}). 
\eea
One notes that both the theoretical results, (\ref{shsa}) and
(\ref{mhsa}), fail by at least an order of magnitude. To assess the
uncertainties of the theoretical results one may vary the parameters,
$m_c$, $B_2$ and $\Lambda_{QCD}$. However, even if the parameters are
pushed to their extreme values the predicted rates are well below
data. Thus, one has to conclude that calculations based on the
assumption that the $\chi_{cJ}$ is a pure $c\bar{c}$ state, are not
sufficient to explain the observed rates. The necessary corrections
would have to be larger than the leading terms. A new mechanism is
therefore called for. 

Recently, the importance of higher Fock states in understanding the
production and the {\em inclusive} decays of charmonium has been
pointed out \cite{bod:95}. It is therefore tempting to assume the inclusion of
contributions from the $|c\bar{c}_8 (^3S_1)g\rangle$ Fock state to
{\em exclusive} $\chi_{cJ}$ decays as the solution to the failure of
the HSA. The usual higher Fock state suppression by powers of $1/Q^2$
\cite{far:73} where $Q=m_c$ in the present case, does not appear as a
simple dimensional argument reveals: the colour-singlet and octet 
contributions to the decay amplitude behave as
\be
  \label{powers}
  M^{(c)}_J \sim f_{\pi}^2 f^{(c)}_J m_c^{-n_{c}}.
\ee
The singlet decay constant, $f^{(1)}_J$, represents the derivative of a
two-particle coordinate space wave function at the origin. Hence it is
of dimension GeV$^2$. The octet decay constant, $f^{(8)}_J$, as
a three-particle coordinate space wave function at the origin, is also
of dimension GeV$^2$. Since $M^{(c)}_J$ is of dimension GeV, $n_{c}=3$
in both cases. Note that the $\chi_{c J}$ decay constants may also 
depend on $m_c$. Obviously, the colour-octet contribution will also
play an important role in the case of the $\chi_{b J}$ decays.

\begin{figure}[t]
\unitlength 1mm
 \begin{picture}(160,100)
   \put( 5,30){\psfig{figure=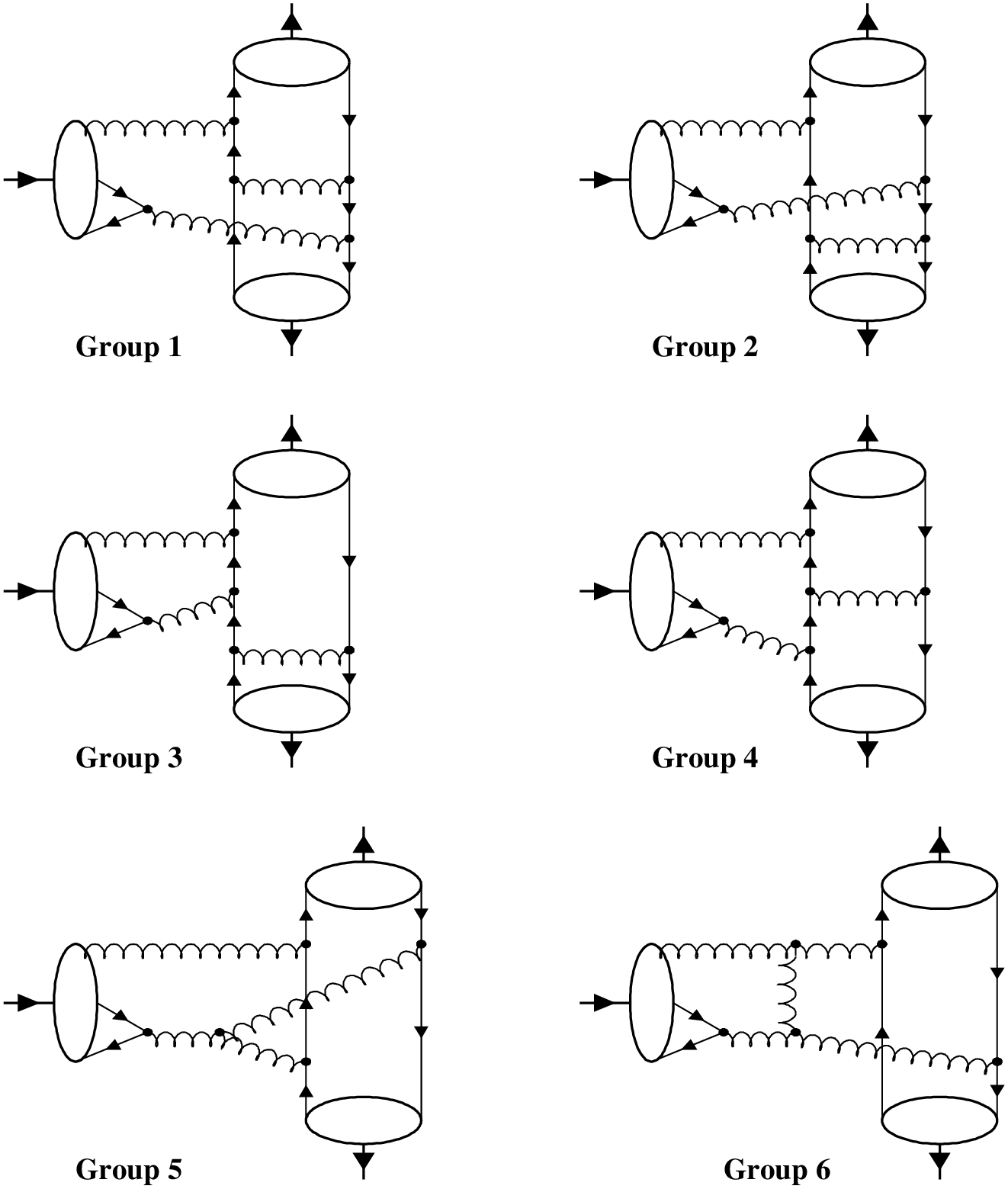,%
       bbllx=10pt,bblly=335pt,bburx=580pt,bbury=810pt,%
       width=6.8cm,clip=} }
   \put(80,30){\psfig{figure=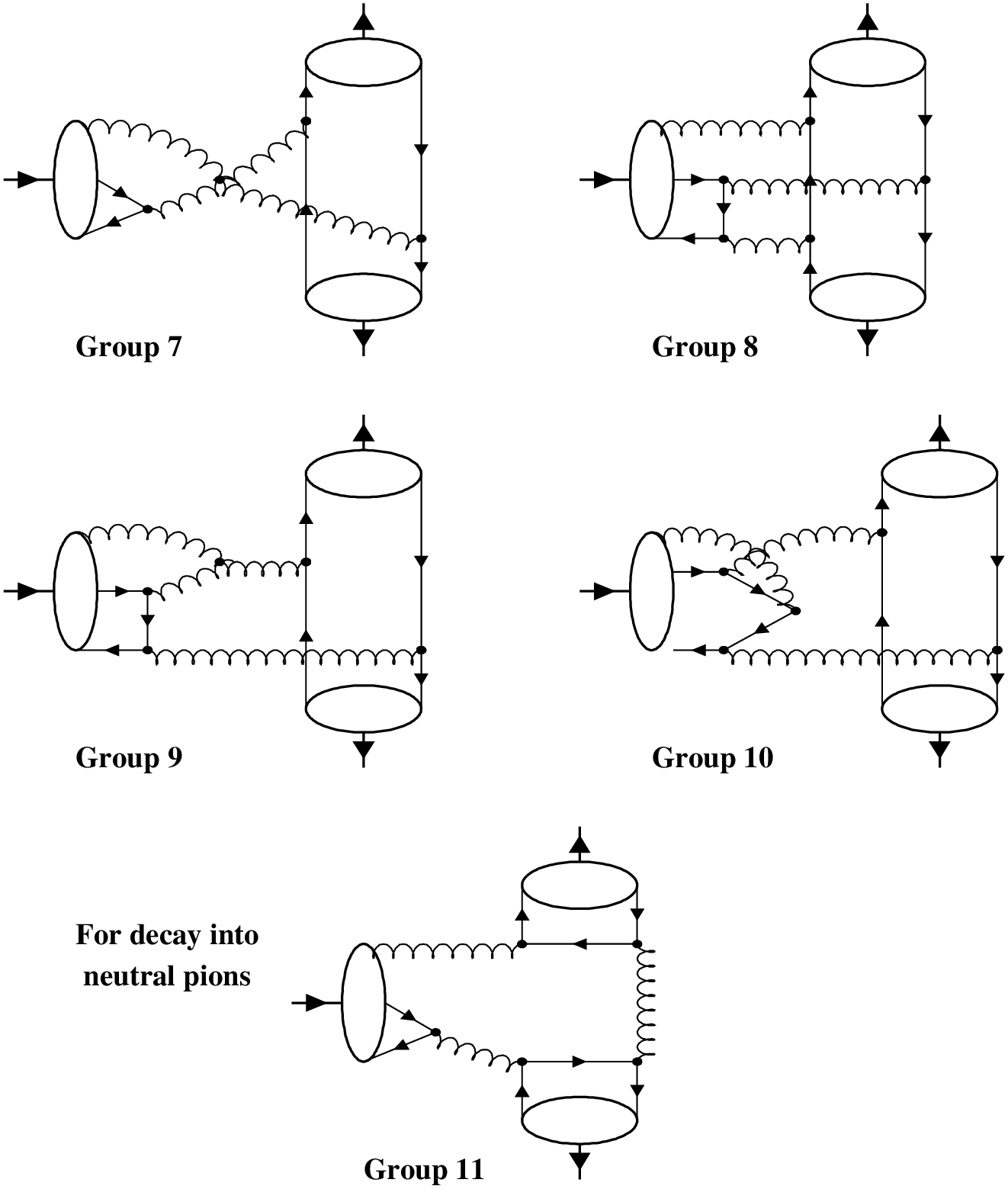,%
       bbllx=10pt,bblly=335pt,bburx=580pt,bbury=810pt,%
       width=6.8cm,clip=} }
 \end{picture}
%\framebox[70mm]{\rule[-21mm]{0mm}{43mm}}
\caption{Representatives of the various groups of colour-octet decay graphs.
 \label{fig:col}}
 \vspace{-0.3cm}
\end{figure}

In \cite{bol:96} the colour-octet contributions to the
exclusive $\chi_{cJ}$ decays are estimated by calculating the hard
scattering amplitude from the set of Feynman graphs shown in Fig.\
\ref{fig:col} and convoluting it with the asymptotic pion wave function.
The colour-octet and singlet contributions are to be added coherently. The
$\chi_{cJ}\to \pi\pi$ decay widths are given in terms of a single
non-perturbative parameter $\kappa$ which approximately accounts for
the soft physics in the colour-octet contribution. A fit to the data
\cite{pdg,bes} yields $\kappa=0.16 \gev^2$ (with $m_c=1.5$ GeV; 
$\Lambda_{QCD}=0.2$ GeV) and the widths
\be
\label{so}
\Gamma (\chi_{c0(2)}\to\pi^+\pi^-)\, =\, 49.85\; (3.54)\,  \kev. 
\ee
Comparison with (\ref{dat}) reveals that the inclusion of the
colour-octet mechanism brings predictions and data in generally good
agreement. The value found for the parameter $\kappa$ has a reasonable
interpretation in terms of charmonium properties and the mean transverse 
momentum of the quarks inside the pions. Results for the decays into 
pairs of uncharged pions are presented in Table 2. The quoted
results for the pionic charmonium decays refer to a calculation within the
standard HSA but similar good results are found when the modified HSA is
used \cite{bol:97}. The only soft parameter appearing in the latter
calculation is the octet-decay constant $f_J^{(8)}$ of the charmonium state.
\begin{table}
\vspace*{0.5cm}
\begin{center}
\begin{tabular}{|c|r|r|r|r|} \hline
  & \multicolumn{2}{|c|}{$\Gamma(\chi_{c J} \to \pi^0\pi^0)\,$[keV]  }
  & \multicolumn{2}{|c|}{${\rm BR}(\chi_{c J} \to \pi^0\pi^0)\,$[\%]}
\\ \hline
  $B_2 = 0$ & $25.7$ & $1.81$ & $0.18$ & $0.091$
\\ \hline
 Exp.\ PDG \cite{pdg} & $42 \pm 18$ & $2.2 \pm 0.6$ & 
  $0.31 \pm 0.06$ & $0.110 \pm 0.028$
\\ \hline
\end{tabular}
\caption[dummy]{Decay widths and branching ratios of $\chi_{c J}
\rightarrow \pi^0 \pi^0$ (colour-octet contributions included; 
$m_c=1.5\,$GeV, $\Lambda_{QCD}=0.2\,$GeV).}
\label{tab:pionzero}
\end{center}
\end{table}

Thus it seems that the colour-octet mechanism leads to a
satisfactory explanation of the decay rates of the $\chi_{cJ}$ into 
two pions. Of course, that mechanism has to pass more tests in
exclusive reactions before this issue can be considered as being
settled. 

%%%%%%%%%%%%%%%%%%%%%%%%%%%%%%%%%%%%%%%%%%%%%%%%%%%%%%%%%%%%%%%%%%%%%%%%%% 
\section{The $\pi$ form factor}
\label{sec:pion ff}
%%%%%%%%%%%%%%%%%%%%%%%%%%%%%%%%%%%%%%%%%%%%%%%%%%%%%%%%%%%%%%%%%%%%%%%%%%
\vspace*{-0.3cm}
Let us now turn to the case of the $\pi$ form factor and discuss the 
implications of the constraints on the pion's wave function obtained from
the $\pi\gamma$ analysis. The leading twist result for the pion form
factor can be brought into a form similar to (\ref{leadtwisteq})
\be
\label{leadtwist}
F_{\pi}(Q^2) = \frac{8\pi}{9}\,\langle
                 x^{-1}\rangle^2 \frac{f_\pi^2}{Q^2}\;\al(\mu_R)\;
             [\,1+\frac{\alpha_s(\mu_R)}{2\pi}
               K_{\pi}(Q^2,\mu_R) + {\cal O}(\alpha_s^2)\,]. 
\ee
Choosing $\mu_F=\mu_R$ and using the value of $B_2^{LO}(\mu_0)$
determined in the leading twist analysis of the $F_{\pi\gamma}$ as
well as a value of, say, 0.4 for $\al$, one obtains the result 
$0.097 \gev^2/Q^2$ for $F_{\pi}$ to lowest order pQCD. That result is 
much smaller than the admittedly poor experimental result \cite{beb:76}: 
$F_{\pi}=0.35\pm 0.10\gev^2/Q^2$. The $\al$-corrections are too small 
to account for that discrepancy \cite{fie:81}. The modified HSA likewise 
provides too small a perturbative contribution \cite{jak:93}. 
It is important to remember at this point that, formally, the perturbative 
contribution to the pion form factor represents the overlap of the large 
momentum tails of the initial and final state wave functions. But 
the form factor also gets a contribution from the overlap of the 
soft wave functions $\hat\Psi_{\pi}$. That contribution, frequently 
termed the Feynman contribution, is customarily assumed to
be negligible already at momentum transfers as low as a few GeV$^2$\footnote{
At large $Q^2$ the Feynman contribution is suppressed by $1/Q^2$ as compared 
to the perturbative contribution.}. 
Examining the validity of that presumption by estimating the Feynman 
contribution from the asymptotic wave function (\ref{gaussian}), one 
finds results of appropriate magnitude to fill in the gap between the 
perturbative contribution and the data of \cite{beb:76}. The results 
exhibit a broad flat maximum which, for momentum transfers between 
$3$ and about $15$ GeV$^2$, simulates the dimensional counting behaviour. 
For a wave function based on the Chernyak-Zhitnitsky \da, on the other
hand, the Feynman contribution exceeds the data significantly 
\cite{jak:96,jak:93}. Large Feynman contributions have also been 
found by other authors \cite{isg:89,KisWan:93}.  
Thus, the small size of the perturbative contribution 
to the elastic form factor finds a comforting although model-dependent
explanation, a fact which has been pointed out by Isgur and 
Llewellyn Smith \cite{isg:89} long time ago. Of course this line of 
reasoning is based on the assumption that the data of \cite{beb:76}
are essentially correct.

A comment concerning the pion form factor in the time-like region is
in order. In the standard HSA the predictions for the form factor in
both the time-like and the space-like region, are
identical. The experimental information on the time-like form
factor comes from two sources, 
$e^+\,e^-\to \pi^+\pi^-$ and $J/\Psi \to\pi^+\pi^-$, 
which provides, to a very good approximation\footnote{
The contribution from $c\bar{c}$ transitions into the light quarks via
three gluons cancel to zero if quark masses are neglected \cite{BL}.},
the form factor at $s=M^2_{J/\Psi}$. Although the 
$e^+\,e^-$ annihilation data of Bollini et al.\ \cite{Bol75}
suffer from low statistics, they agree very well with the result obtained
from the $J/\Psi$ decay. Combining both the data, one finds
$|F_{\pi}|=0.93\pm 0.08\gev^2/s$ in the momentum transfer range
between 2 and 10 GeV$^2$, a value wich is roughly a factor of
3 larger than the space-like data. The modified HSA can account for that
large ratio of the time-like over space-like form factors \cite{GP}
although the perturbative contributions to the form factor in both
the regions are too small.

The structure function of the pion offers another possibility to test the wave 
function against data. As has been pointed out in \cite{BHL:83} the parton 
distribution functions are determined by the Fock state wave functions. Since 
each Fock state contributes through the modulus squared of its wave function 
integrated over transverse momenta up to $Q$ and over all fractions $x$ except 
those pertaining to the type of parton considered, the contribution from the 
valence Fock state should not exceed the data of the valence quark structure 
function. As discussed in \cite{jak:96,HuaMaShe:94} the asymptotic wave 
function respects this inequality while the Chernyak-Zhitnitsky one again 
fails dramatically.
\newpage
%%%%%%%%%%%%%%%%%%%%%%%%%%%%%%%%%%%%%%%%%%%%%%%%%%%%%%%%%%%%%%%%%%%%%%%%%%%
\section{Summary}
\label{sec:concl}
%%%%%%%%%%%%%%%%%%%%%%%%%%%%%%%%%%%%%%%%%%%%%%%%%%%%%%%%%%%%%%%%%%%%%%%%%%%
\vspace*{-0.3cm}
The study of hard exclusive reactions is an interesting and
challenging subject. The standard HSA, i.e.\ the valence Fock state
contribution in collinear approximation to lowest
order perturbative QCD, while asymptotically correct (at
least for form factors), does not lead to a consistent description of
the data. In many cases the predicted perturbative contribution to
particular exclusive reactions are much smaller then the data. The
observed spin effects do not find a comforting explanation. In some
reactions agreement between prediction and experiment is found
although at the expense of dominant contributions from the soft
end-point regions rendering the perturbative analysis inconsistent.

From a detailed analysis of the $\pi\gamma$ transition form factor it
turned out that the pion \da\ is close to the asymptotic form. Strongly
end-point concentrated \das\ are obsolete and the ostensible successes
in describing the large momentum transfer behaviour of the pion form 
factor in the space-like region and charmonium decays into two pions 
with such \das\ must therefore be dismissed.

In view of these observations it seems that higher Fock state and/or 
higher twist contributions have to be included in the analysis of 
exclusive reactions. However, not much is known about them as yet. 
We are lacking systematic investigations of such contributions to 
exclusive reactions. The colour octet model for exclusive charmonium 
decays is discussed in this talk as an example of such contributions.

%\newpage

\end{document}